\documentclass{pasj00}
\draft

\def\ion#1#2{#1\,{\sc #2}}
\newcommand{\lam}{$\lambda$}
\newcommand{\kms}{km~s$^{-1}$}
\newcommand{\ecs}{erg~cm$^{-2}$~s$^{-1}$~sr$^{-1}$}
\input{epsf}

\begin{document}
\SetRunningHead{P.R. Young et al.}{Solar transition region features observed
  with EIS} 
%\Received{2000/12/31}%{yyyy/mm/dd}
%\Accepted{2001/01/01}%{yyyy/mm/dd}

\title{Solar transition region features observed with Hinode/EIS}

%%% begin:list of authors
% Do NOT capitalize all letters in "textsc".

\author{P.R. \textsc{Young},\altaffilmark{1}
        G. \textsc{Del Zanna},\altaffilmark{2}
        H.E. \textsc{Mason},\altaffilmark{3}
        G.A. {\sc Doschek},\altaffilmark{4}
        J.L. {\sc Culhane},\altaffilmark{2}
        H. \sc{Hara}\altaffilmark{5}}
\altaffiltext{1}{STFC Rutherford Appleton Laboratory, Chilton, Didcot,
  Oxfordshire, OX11 0QX, U.K.}
\altaffiltext{2}{University College London, Department of Space and Climate
  Physics, Holmbury St. Mary,
  Dorking, Surrey, UK}
\altaffiltext{3}{DAMTP, Centre for Mathematical Sciences, Cambridge,
  UK}
\altaffiltext{4}{Code 7670, Naval
  Research Laboratory, Washington, DC 20375-5352,
  USA}
\altaffiltext{5}{National Astronomical Observatory of Japan, National Institutes
  of Natural Sciences, Mitaka, Tokyo 181-8588}

%%% end:list of authors

%%% Please use the following style in case that sorting by 
%%% affilation is impossible. 
%
% \author{%
%   D-Firstname \textsc{D-Familyname}\altaffilmark{1}
%   E-Firstname \textsc{E-Familyname}\altaffilmark{1,2}
%   and
%   F-Firstname \textsc{F-Familyname}\altaffilmark{2}}
% \altaffiltext{1}{Address of Institute}
% \email{ddddd@xxx.xxx.xx.xx}
% \email{eeeee@xxx.xxx.xx.xx}
% \altaffiltext{2}{Address of Institute}

%% `\KeyWords{}' always has to be placed before `\maketitle'.
\KeyWords{Sun: transition region -- Sun: UV radiation} %Do NOT move this preamble from here!

\maketitle

\begin{abstract}
Two types of solar active region feature prominent at transition region
temperatures are identified in Hinode/EIS data
of AR 10938 taken on 2007 January 20. The footpoints of 1~MK TRACE
loops are shown to emit strongly in emission lines formed at
$\log~T=5.4$--5.8, allowing the temperature increase along the
footpoints to be clearly seen. A density diagnostic of \ion{Mg}{vii}
yields the density in the footpoints, with one loop showing a decrease
from $3\times 10^9$~cm$^{-3}$ at the base to $1.5\times
10^9$~cm$^{-3}$ at a projected height of 20~Mm. The second feature is
a compact
active region transition region brightening which is particularly
intense in \ion{O}{v} emission ($\log~T=5.4$) but also has a signature
at temperatures up to $\log~T=6.3$. The \ion{Mg}{vii} diagnostic gives
a density of $4\times 10^{10}$~cm$^{-3}$, and emission lines of
\ion{Mg}{vi} and \ion{Mg}{vii} show line profiles broadened by
50~\kms\ and wings extending beyond $\pm 200$~\kms. Continuum emission
in the short wavelength band is also found to be enhanced, and is
suggested to be free-bound emission from recombination onto He$^+$.
\end{abstract}

\section{Introduction}

The Extreme ultraviolet Imaging Spectrometer (EIS) instrument on Hinode \citep{culhane07,kosugi07} covers the two
wavelength bands 170--211 and 
246--292~\AA\ that are dominated by coronal emission lines mainly from the
iron ions. The majority of strong transition region lines in the solar
spectrum are found at
longer UV wavelengths, and all of the transition region lines found in
the EIS bands are weak in normal conditions. \citet{young07} however
identified two types of active region features that were found from
SOHO/CDS observations to yield significantly enhanced transition
region emission lines: coronal loop footpoints and active region
blinkers. He suggested that in such events the weak EIS 
lines would become significant, and so yield useful science. An EIS
observation from 2007 January 20 is presented here that displays both
types of active region feature identified by \citet{young07} and 
demonstrates the value of including transition region lines in EIS
studies.

\section{Data}

\begin{figure*}[h]
\centerline{\epsfxsize=17.5cm\epsfbox{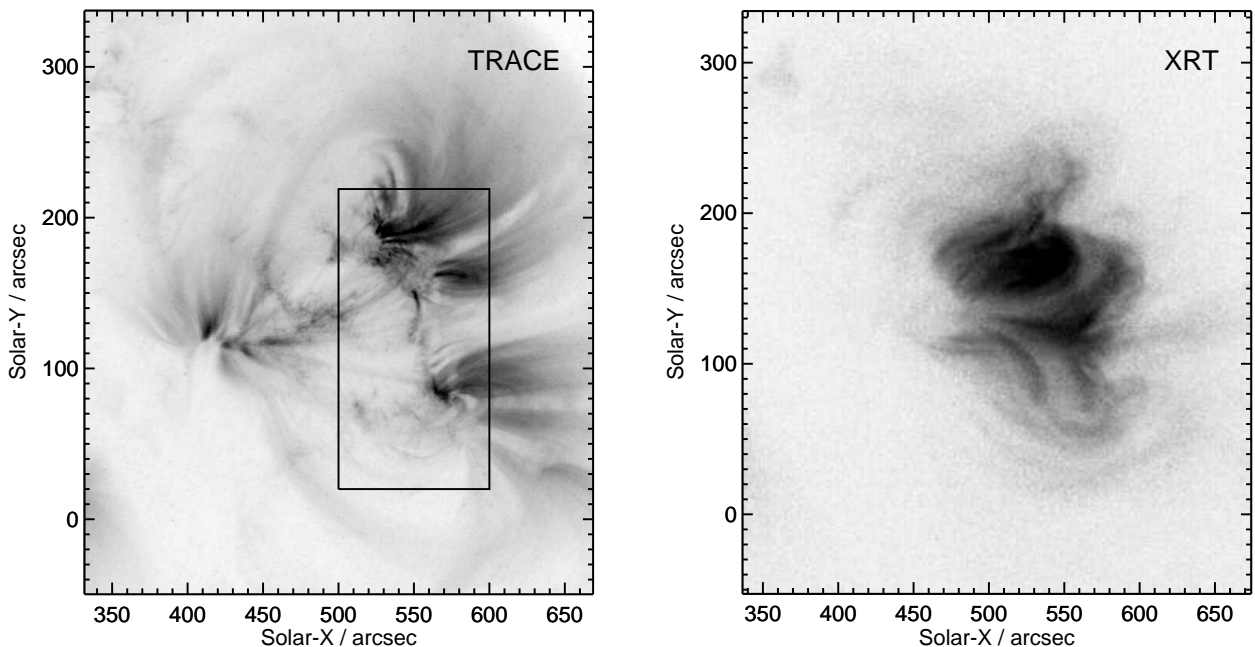}}
\caption{The left-hand panel shows a TRACE 171 filter image taken at
  22:40~UT. The approximate location of the EIS raster is indicated by a
  black box. The right-hand panel shows a XRT image taken with the
  Al-Poly filter at 22:36~UT. For both images dark areas are bright
  in intensity.}
\label{fig.trace-xrt}
\end{figure*}

Active region AR 10938 crossed the solar disk during 2007 January
12--24. It was a well developed active region showing large cool loop
structures visible in the TRACE 171 channel
(Fig.~\ref{fig.trace-xrt}, left panel), and a more compact high temperature core
visible in Hinode/XRT images (Fig.~\ref{fig.trace-xrt}, right panel)
 During the
period January 18--23 the EIS observing study PRY\_loop\_footpoints
was run a number of times at the footpoint regions of the loop
system. This study uses the 2\arcsec\ slit to raster over an area of
100\arcsec $\times$216\arcsec\ with 30 second exposure times, giving a total
duration of 26~minutes. The 2\arcsec\ slit was used to both increase
the counts for weak lines, and allow a larger spatial area to be
covered more rapidly; the disadvantages are a degradation of the
spatial (in the X-direction) and spectral
resolution over the 1\arcsec\ slit.
Due to on board data storage restrictions,
only a fraction of the total EIS wavelength range could be downloaded,
and so 20 wavelength windows were chosen to observe the transition
region lines as well as a number of coronal lines. The list of
the transition region lines (defined as emission lines whose
temperature of maximum ionization, $T_{\rm max}$, is below
$\log\,T=5.8$) is given in Table~\ref{tbl.lines}. Further discussion
of useful emission transition region lines observed with EIS is given
in \citet{young07b}.

\begin{table*}[h]
\caption{Transition region emission lines observed with
    the PRY\_loop\_footpoints EIS study.}
\smallskip
\begin{center}
\begin{tabular}{llll}
\hline
\noalign{\smallskip}
Ion &Wavelength/\AA &Transition &Log\,$T_{\rm max}$/K\\
\hline
\noalign{\smallskip}
\ion{O}{v} & 192.8$^a$
    &2s2p $^3P^{\rm o}_{0,1}$ -- 2s3d $^3$D$_{1,2}$ & 5.4 \\
\ion{O}{v} & 192.9$^b$
    &2s2p $^3P^{\rm o}_2$ -- 2s3d $^3$D$_{3,2}$ & 5.4 \\
\ion{Mg}{v} &276.58 
    &2s$^2$2p$^4$ $^1$D$_2$ -- 2s2p$^5$ $^1P^{\rm o}_1$ &5.4 \\
%\ion{O}{vi} & 184.12$^c$ & $2p$ $^2P_{1/2}$ -- $3s$ $^2S_{1/2}$ &5.5\\
\ion{Fe}{viii} &185.12 
    &3p$^6$3d $^2$D$_{5/2}$ -- 3p$^5$3d$^2$ $^1P^{\rm o}_{1}$ &5.6\\
\ion{Mg}{vi} &268.99
    &2s$^2$2p$^3$ $^2D^{\rm o}_{3/2}$ -- 2s2p$^4$ $^2$P$_{1/2}$ &5.6\\
&270.40
    &2s$^2$2p$^3$ $^2D^{\rm o}_{3/2,5/2}$ -- 2s2p$^4$ $^2$P$_{3/2}$ &5.7\\
\ion{Mg}{vii} &278.39
    &2s$^2$2p$^2$ $^3$P$_2$ -- 2s2p$^3$ $^3S^{\rm o}_1$ & 5.8 \\
&280.75
    &2s$^2$2p$^2$ $^1$D$_2$ -- 2s2p$^3$ $^1P^{\rm o}_1$ & 5.8 \\
\ion{Si}{vii} &275.35
    &2s$^2$2p$^4$ $^3$P$_2$ -- 2s2p$^5$ $^3P^{\rm o}_2$ &5.8 \\
\hline
\noalign{\smallskip}
\multicolumn{4}{l}{$^a$ comprises three lines with wavelengths
  192.750, 192.797, 192.801~\AA.}\\
\multicolumn{4}{l}{$^b$ comprises two lines with wavelengths 192.904
  and 192.911~\AA.}\\
%\multicolumn{4}{l}{$^c$ not included in present study.}\\
\end{tabular}
\end{center}
\label{tbl.lines}
\end{table*}

\begin{figure*}[h]
\centerline{\epsfxsize=17.5cm\epsfbox{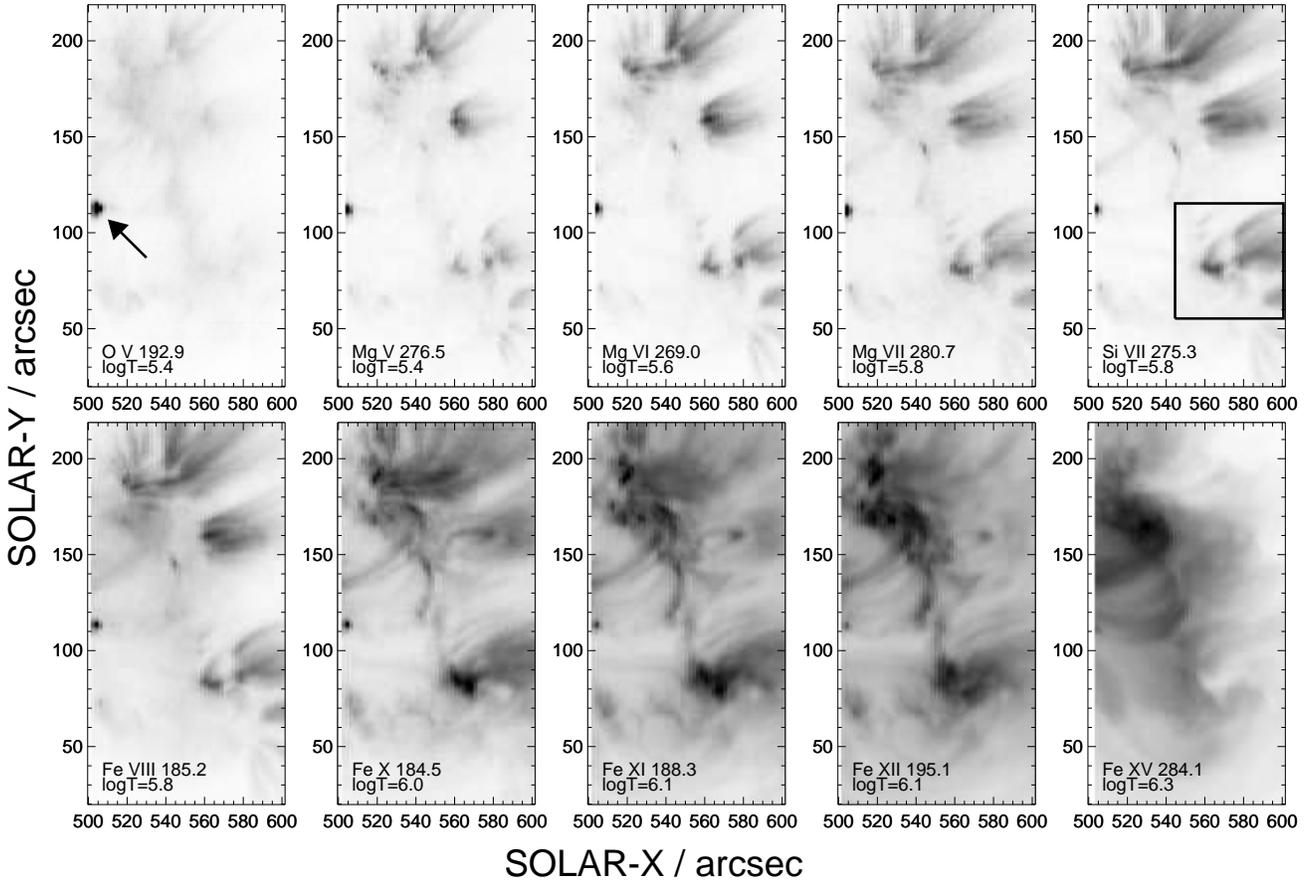}}
\caption{EIS raster images in ten different emission lines for the
  spatial region indicated in the TRACE image of
  Fig.~\ref{fig.trace-xrt}. The loop footpoints are most clearly seen
  in \ion{Mg}{v--vii}, \ion{Si}{vii} and \ion{Fe}{viii}. The box on
  the \ion{Si}{vii} image indicates the region shown in
  Fig.~\ref{fig.footpoint-closeup}, while the arrow in the \ion{O}{v}
  image points to the location of the ARTRB.}
\label{fig.ion-images}
\end{figure*}

Raster images for a number of the emission lines are shown in
Fig.~\ref{fig.ion-images}. The two EIS wavelength bands are imaged
onto different CCDs and there is a spatial offset in the Solar-Y
direction between the CCDs that varies with wavelength and ranges from
16 to 20 pixels depending on the wavelengths being compared. A fixed
offset 
of 17 pixels has been applied to the images in
Fig.~\ref{fig.ion-images}. In addition, when comparing raster images
from the two CCDs, there is a spatial offset in
the solar-X direction of around 2~arcsec. The images have also been
corrected for this effect.

The data have been calibrated to yield intensities in units of \ecs\
at each pixel in the data-set. The dark current and CCD pedestal were
removed by subtracting the median value of the darkest 2~\% pixels in
each wavelength window. Cosmic rays and CCD hot pixels were removed
using the IDL routine NEW\_SPIKE. The conversions from data numbers to
photons and on to calibrated intensites were performed using data
contained in the EIS directory of the \emph{Solarsoft} IDL
distribution.

\section{Loop footpoints}

Fig.~\ref{fig.ion-images} shows a number of `clumps' of loop
footpoints that can be readily identified with the loops in the TRACE
171 image (Fig.~\ref{fig.trace-xrt}). The spectroscopic properties of
such loops have been studied previously by \citet{delzanna03a} and
\citet{delzanna03b} using SOHO/CDS spectra. In particular, there is a
steep temperature increase at the loop base, leading to transition
region lines being formed in a small spatial area at the footpoint of
the loop. The main body of the loop is at temperatures of around 1~MK
and thus strongly emits in the \ion{Fe}{ix} \lam171.1 and \ion{Fe}{x}
\lam174.5 emission lines that contribute to the TRACE 171
passband. The \ion{Mg}{vii} \lam319.0/\lam367.7 density diagnostic 
yielded densities of $\approx$ $2\times 10^9$~cm$^{-3}$.

Although EIS lacks the strong transition region lines (such as
\ion{Ne}{vi} \lam562.8, \ion{O}{v} \lam629.7) observed by SOHO/CDS,
the images in Fig.~\ref{fig.ion-images} clearly demonstrate that the
footpoints of the 1~MK loops can be identified in the EIS data.
In particular, enhanced emission at the base of
the loops can 
be seen in \ion{Mg}{v}, \ion{Mg}{vi} and, to a lesser extent,
\ion{O}{v}. A comparison from spectra in and outside of the footpoint
regions is shown in Fig.~\ref{fig.footpoint-spectra}, demonstrating
the strong enhancement of the cool lines. Generally the cool Mg
emission can be identified to extend 
for a significant length along the loops (10--50~arcsec). We
demonstrate below how the EIS spectra can be used to derive temperature,
density and 
filling factor information about these loops.

\subsection{Temperature analysis}

A simply visual inspection of the loop footpoint images in
Fig.~\ref{fig.ion-images} shows that the
\ion{Fe}{viii} \lam185.21 structures look very similar to
\ion{Si}{vii} \lam275.35 structures. This is at odds with the
temperatures of maxium ionization given in Table~\ref{tbl.lines} that
are derived from the \citet{mazz98} ion balance calculations. The
ionization and recombination rates are expected to be considerably
more uncertain for Fe$^{7+}$ than those for Si$^{6+}$ given 
its more complex atomic structure. A conclusion from the EIS images is
thus that 
\ion{Fe}{viii} is actually principally formed at a temperature of
$\log\,T=5.8$. This will have a knock-on effect for \ion{Fe}{ix} which
also must be formed at a higher temperature than given by
\citet{mazz98}. 

The different appearance of the loop footpoint regions in the
\ion{Mg}{v}, \ion{Mg}{vi} and \ion{Si}{vii}  lines clearly
demonstrates that the loops are not isothermal in the lowest $\approx
10$~Mm of the loop. We defer to a later paper a discussion of whether
the loops are 
actually multithermal in 
this region (i.e., a range of temperatures applies at a given point in
the loop), or whether the temperature is simply decreasing towards the
footpoints. \citet{delzanna03a} inferred the latter from his analysis
of SOHO/CDS spectra.

%% Whether the loops are actually multithermal in
%% this region (i.e., a range of temperatures applies at a given point in
%% the loop), or whether the temperature is simply decreasing towards the
%% footpoints will require a more detailed analysis.

\begin{figure}[h]
\centerline{\epsfxsize=8.5cm\epsfbox{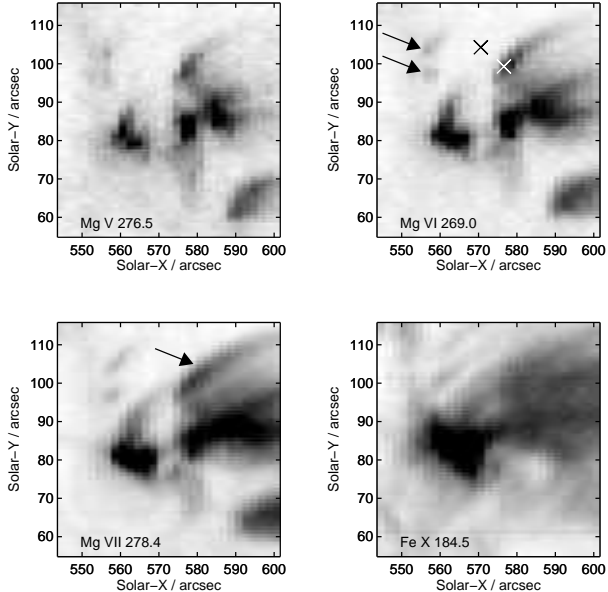}}
\caption{A close-up of the footpoint region indicated in the
  \ion{Si}{vii} image of Fig.~\ref{fig.ion-images}. Arrows point to
  two weak footpoints in the \ion{Mg}{vi} image that are the
  compact footpoints of hot loops. The arrow in the \ion{Mg}{vii}
  image points to the loop selected for density analysis. The two
  crosses on the \ion{Mg}{vi} image indicate the spatial pixels from
  which the spectra shown in Fig.~\ref{fig.footpoint-spectra} were
  obtained. }
\label{fig.footpoint-closeup}
\end{figure}

\begin{figure*}[h]
\centerline{\epsfxsize=17.5cm\epsfbox{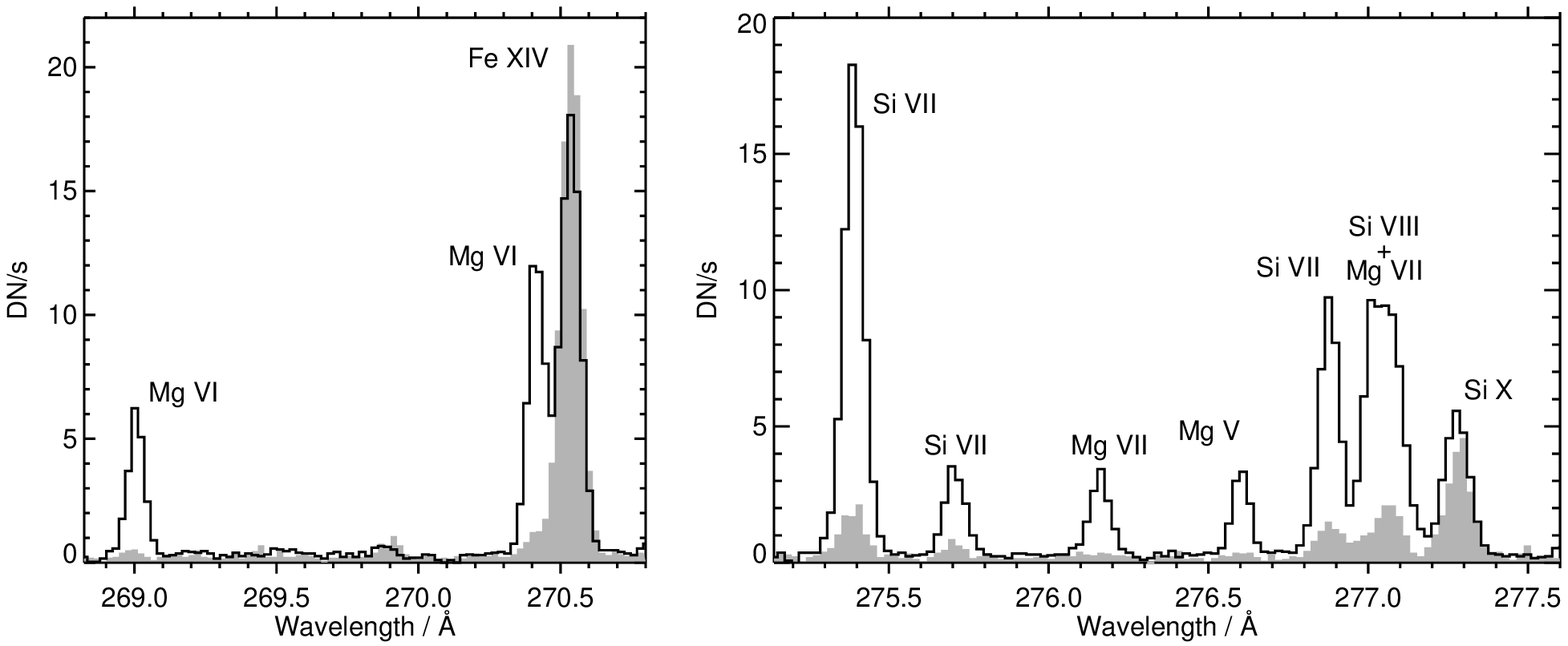}}
\caption{A comparison of spectra at the two pixels identified in
  Fig.~\ref{fig.footpoint-closeup}. The spectrum from within the
  footpoint (white cross in Fig.~\ref{fig.footpoint-closeup}) is
  identified with a solid line. The shaded spectrum is from a pixel
  outside of the footpoint (black cross in
  Fig.~\ref{fig.footpoint-closeup}). The Y-axis shows the strength of
  the lines in data numbers (DN) per second.}
\label{fig.footpoint-spectra}
\end{figure*}

Two weak footpoints are highlighted by two arrows in the \ion{Mg}{vi} image of
Fig.~\ref{fig.footpoint-closeup}. They show a different character to
the other more extended footpoints in the image, being
significantly more compact in \ion{Mg}{v--vii}. Comparing the
different images in Fig.~\ref{fig.ion-images} and also the TRACE image
(Fig.~\ref{fig.trace-xrt}), shows that these loops are hotter, emitting
principally in \ion{Fe}{xii} \lam195.1 rather than \ion{Fe}{ix} and
\ion{Fe}{x}. 

\subsection{Density analysis}

\begin{figure}[h]
\centerline{\epsfxsize=8.5cm\epsfbox{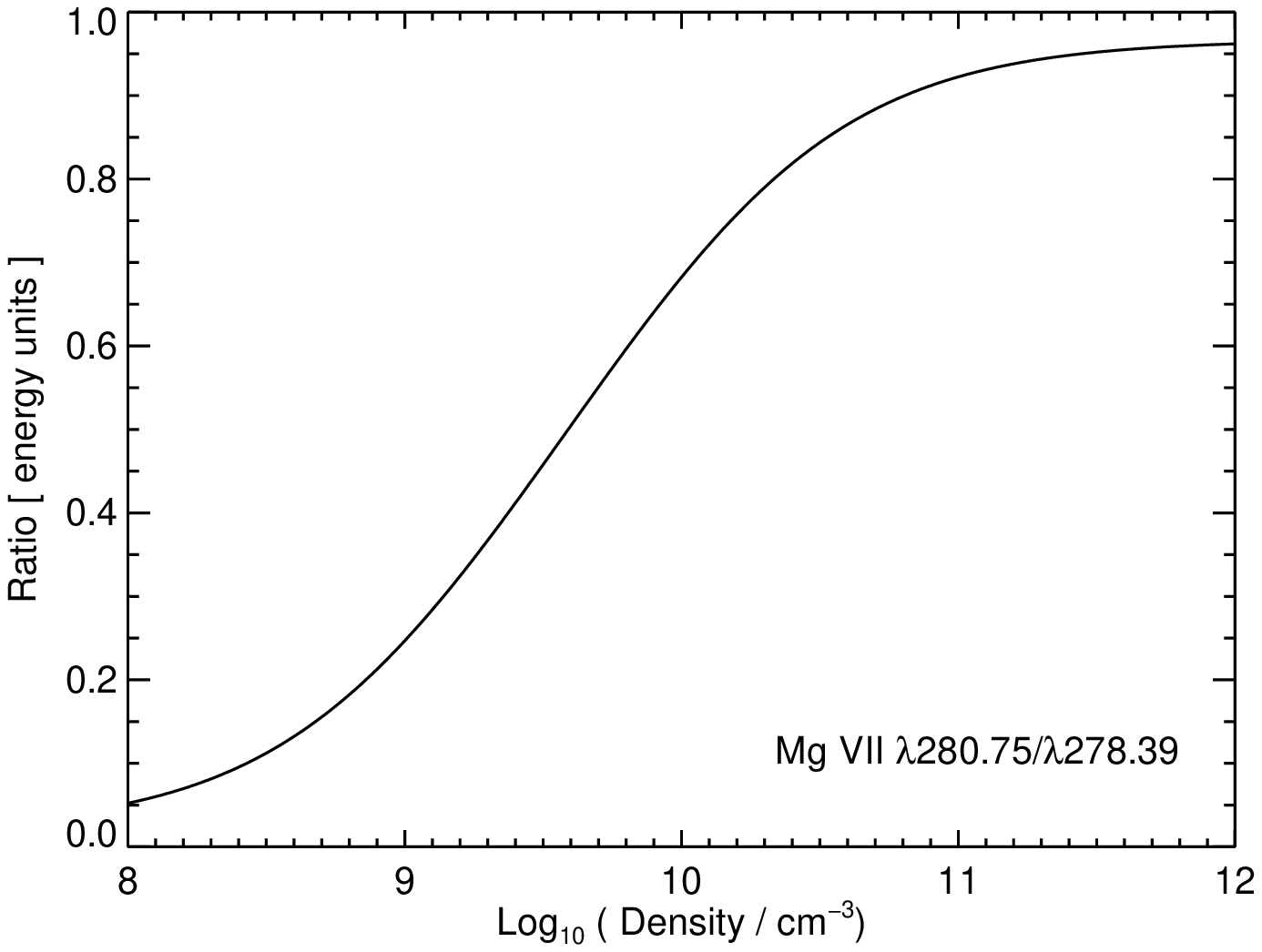}}
\caption{Theoretical variation of the \ion{Mg}{vii}
  \lam280.75/\lam278.39 ratio with density. Produced using v.5.2 of
  the CHIANTI atomic database \citep{landi06}.}
\label{fig.mg7-ratio}
\end{figure}

The EIS long wavelength channel contains the \ion{Mg}{vii}
\lam280.75/\lam278.39  density diagnostic which is directly comparable
to the \lam319.0/\lam367.7 diagnostic that is observed with SOHO/CDS
\citep{young97,delzanna03a}.
The diagnostic is
sensitive to densities in the range $10^8$--$10^{11}$~cm$^{-3}$
(Fig.~\ref{fig.mg7-ratio}), but analysis is complicated by a
blend of the 
\lam278.39 line with \ion{Si}{vii} \lam278.44. A simultaneous two
Gaussian fit to 
the feature is, however, able to resolve the two components
if both are assumed to have the same width (this is a reasonable
assumption since the two ions are formed at very similar
temperatures). Tests of the validity of the two Gaussian fits have
been confirmed with the current data-set by comparing the derived
intensities with the \ion{Si}{vii} \lam275.35 and \ion{Mg}{vii}
\lam276.14 lines. The \ion{Si}{vii} \lam278.44/\lam275.35 and
\ion{Mg}{vii} \lam276.14/\lam278.39 ratios are branching ratios with
fixed theoretical ratios of 0.32 and 0.20, respectively. 

\begin{figure}[h]
\centerline{\epsfxsize=8.5cm\epsfbox{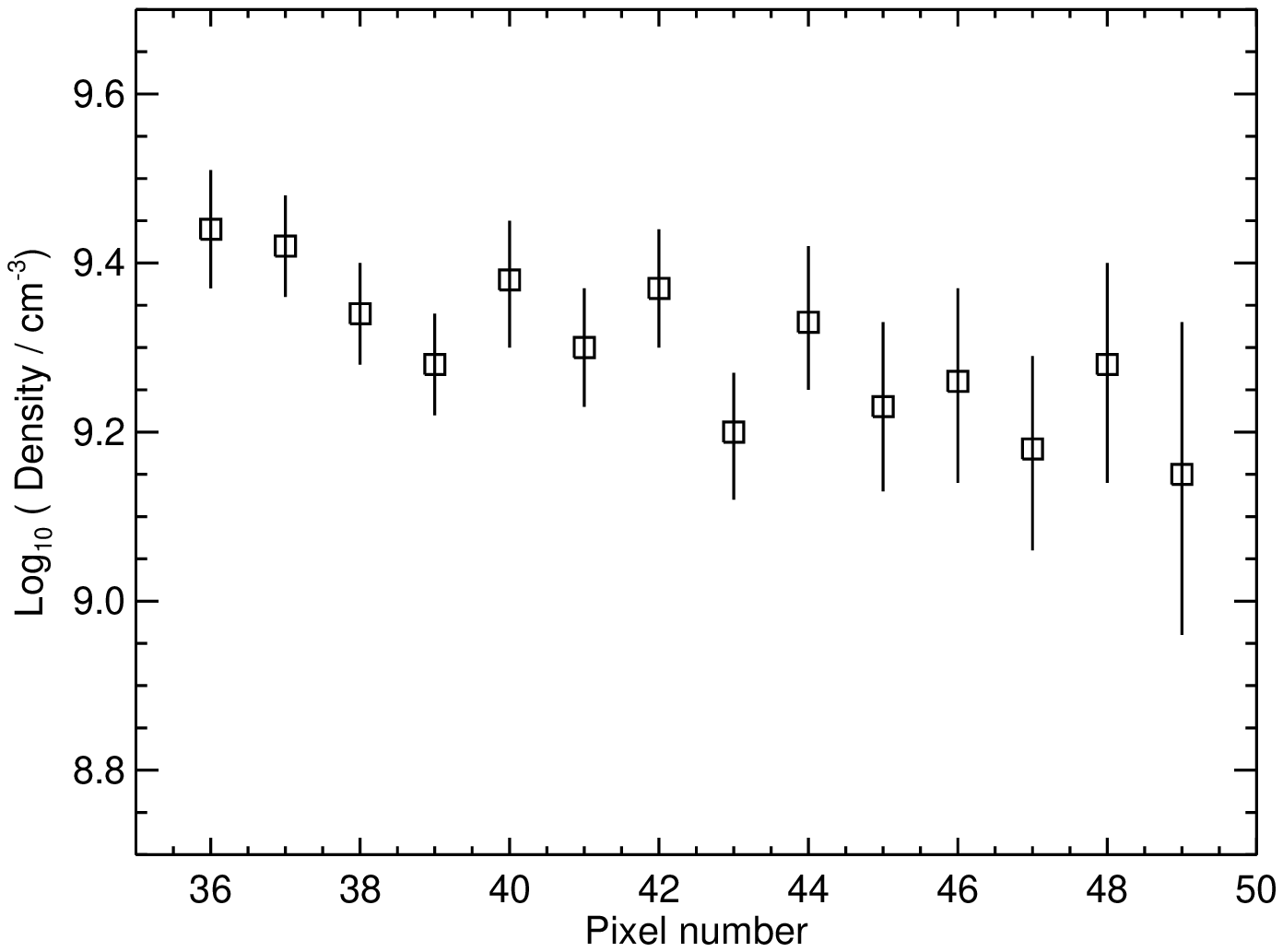}}
\caption{Electron densities derived from the \ion{Mg}{vii}
  \lam280.75/\lam278.39 density diagnostic.}
\label{fig.mg7-densities}
\end{figure}

Fig.~\ref{fig.mg7-densities} shows the densities derived from the
\ion{Mg}{vii} diagnostic for a number of positions along the loop
structure highlighted by an arrow in the \ion{Mg}{vii} image of
Fig.~\ref{fig.footpoint-closeup}. Four to five pixels across the loop 
width (i.e., in the solar-Y
direction) were averaged to derive the line intensities. An estimate of
the background was made by summing a number of pixels in a low intensity region
just to the north of the loop. There is a clear trend of decreasing
density along the structure.
The density estimates are in excellent agreement with those obtained
by \citet{delzanna03a} and \citet{delzanna03b} demonstrating that such
values are typical 
for 1~MK loops. The improved spatial resolution of EIS over CDS now
means that the spatial variation of density in the footpoint regions can be
accurately measured giving additional constraints on theoretical loop
models.

By assuming an isothermal plasma whose temperature is the $T_{\rm
max}$ of \ion{Mg}{vii} it is possible to use the density measurements
to estimate the column depth of the plasma in the loop. The atomic
data from v5.2 of CHIANTI \citep{landi06,dere97} were used together with the
coronal abundances of \citet{feldman92} to derive column depths of
1.6--5.4~Mm (2.3--7.4\arcsec). These values are comparable
to the observed diameter of the loop (around 5--10\arcsec) and thus
suggest a filling factor of around unity.

\section{Transition region brightening}

\begin{figure*}[h]
\centerline{\epsfxsize=17.5cm\epsfbox{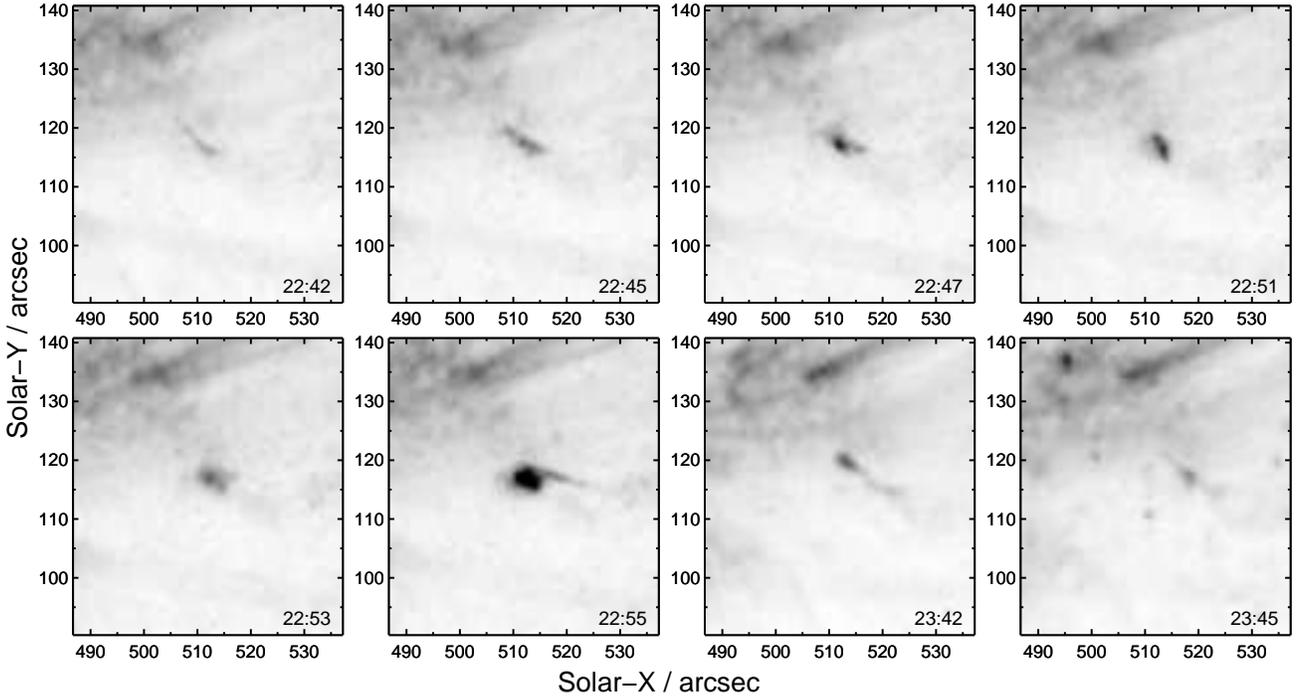}}
\caption{Evolution of the ARTRB as observed by TRACE in the 171
  filter. No TRACE data were available between 22:55 and 23:42 UT. EIS
  observed the ARTRB at 22:57 UT.} 
\label{fig.trace-panel}
\end{figure*}

The small intense brightening (heretoafter referred to as an ARTRB --
active region transition region brightening) seen in the raster images
(Fig.~\ref{fig.ion-images}) was only present in one of the sequence of
EIS rasters. However, the evolution of the event can be seen from
TRACE images (Fig.~\ref{fig.trace-panel}) that demonstrate that it
began at 22:42~UT and had faded 
completely by 23:49~UT. No TRACE data were available between 22:55~UT and
23:42~UT. The peak intensity occurred at 22:55~UT, 
corresponding closely to the time at which EIS scanned the region (22:57~UT). The
brightening is very bright in the transition region at the temperature
of \ion{O}{v} ($\log\,T=5.4$) and the brightness decreases to higher
temperatures, until it is barely discernible in \ion{Fe}{xv}
($\log\,T=6.3$). The brightening is not apparent in co-temporal XRT
images. 

The density and temperature properties (discussed below) appear to be
similar to those 
of impulsive events identified in EUV Skylab spectra by
\citet{widing82}, and to an active region brightening seen in EUV
spectra from SOHO/CDS by \citet{young97}, later termed active region
blinkers and discussed further by \citet{young04}. There are also
similarities in terms of intensity enhancement and line broadening to
the bidirectional jet observed by \citet{doyle04} in TRACE and
SOHO/SUMER data.

\subsection{Oxygen emission lines}

\begin{figure}[h]
\centerline{\epsfxsize=8.5cm\epsfbox{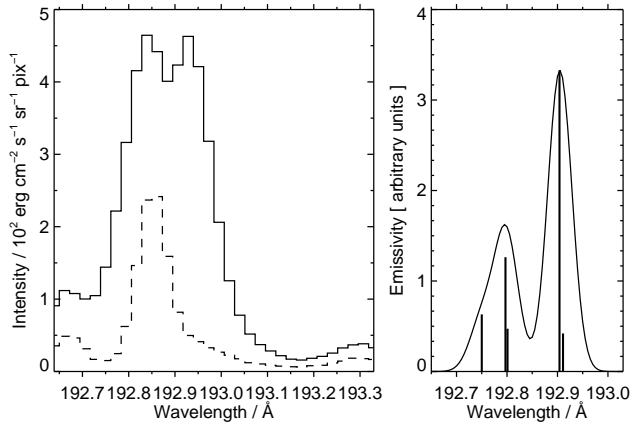}}
\caption{Left panel: a comparison of spectra in the vicinity of the
  \ion{Fe}{xi}--\ion{Ca}{xvii}--\ion{O}{v} blend at
  192.8--192.9~\AA. The solid line is a spectrum from the ARTRB, while
  the dashed line is from a region to the north of 
  the brightening showing the more typical appearance of the
  spectrum. Right panel: a synthetic spectrum generated with CHIANTI
  showing the five \ion{O}{v} components (vertical lines) and the
  resulting spectral distribution (assuming Gaussian line profiles).}
\label{fig.o5-spectra}
\end{figure}

A number of n=3 to 2 transitions of \ion{O}{iv--vi} are found in the
EIS wavebands \citep{young07b}. They are mostly very weak but can be
detected with the 
high spectral resolution and sensitivity of EIS. The \ion{O}{v} 2s2p
$^3P^{\rm o}_J$ -- 2s3d $^3$D$_{J^\prime}$ transitions are of particular
interest as they lie close to \ion{Ca}{xvii} 192.82 -- an EIS core line
that is observed in every EIS study. (This line is actually blended
with \ion{Fe}{xi} \lam192.83 which dominates in most solar
conditions.) The left-hand panel of Fig.~\ref{fig.o5-spectra} compares
spectra of the ARTRB with a region outside of the ARTRB. The dashed
line shows the normally dominant \ion{Fe}{xi} line, while the solid
line demonstrates the strong enhancement of \ion{O}{v} \lam192.9. The
contribution of \ion{O}{v} is complicated as there are actually five
significant \ion{O}{v} components and the CHIANTI prediction for the
relative strengths is shown in the right panel of
Fig.~\ref{fig.o5-spectra}. It is thus seen that \ion{O}{v} makes up
around one half of the emission seen at 192.8~\AA.

%% The right-hand panel of
%% Fig.~\ref{fig.o5-spectra} shows the theoretical prediction from CHIANTI
%% for the  dominant five \ion{O}{v} lines in this
%% region. They divide into two components: one at 192.9~\AA\ is distinct from the
%% \ion{Ca}{xvii}, \ion{Fe}{xi} lines; the other at 192.8~\AA\ is blended
%% with these two lines

%% In the ARTRB discussed here,
%% these \ion{O}{v} lines brighten significantly as can be seen in
%% Fig.~\ref{fig.ion-images}. 

%% Fig.~\ref{fig.o5-spectra} compares the
%% spectrum of the brightening.

%% Because of the blend with both \ion{O}{v} and \ion{Fe}{xi}, studies of
%% the \ion{Ca}{xvii} \lam192.83 line will be difficult

%% Although not observed in the present study, the \ion{O}{vi} \lam184.12
%% line 
%% will be valuable for studying ARTRBs as there is temperature
%% overlap between \ion{O}{vi} and \ion{Mg}{v}, thus potentially allowing
%% estimates of the Mg/O abundance ratio both in ARTRBs and the loop
%% footpoints. 

%% SOHO/CDS spectra have revealed markedly different Mg/Ne
%% abundance ratios in these two solar features \citep{young97}, and a
%% similar result should be expected for Mg/O.

\subsection{Line broadening}

\begin{figure}[h]
\centerline{\epsfxsize=8.5cm\epsfbox{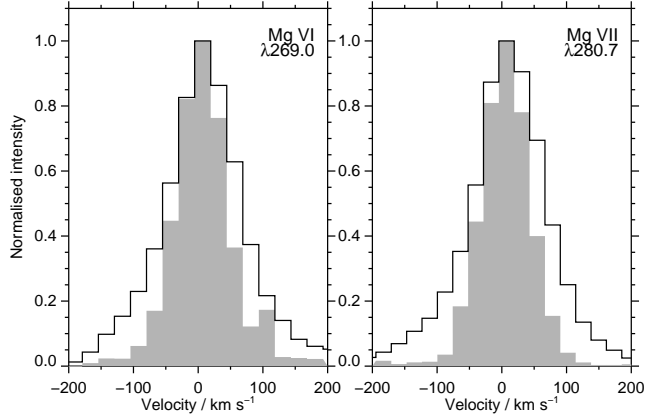}}
\caption{A comparison of emission line profiles for \ion{Mg}{vi}
  \lam269.0 (left-hand panel) and
  \ion{Mg}{vii} \lam280.7 (right-hand panel). The shaded feature gives
  the line profile averaged over a number of spatial pixels in one of
  the loop footpoints, the solid line gives the profile in the ARTRB.}
\label{fig.line-profiles}
\end{figure}

The brightening displays significantly broadened line profiles in the
transition region,  as demonstrated for \ion{Mg}{vi} \lam268.99 and
\ion{Mg}{vii} \lam280.75 in
Fig.~\ref{fig.line-profiles}. The cores of the lines are seen to be
broadened by around 50~\kms\ while the wings extend out beyond $\pm
200$~\kms. Given the strongly dynamic behaviour of the ARTRB in the TRACE
movie, these features are most likely to be due to a superposition of
high speed up- and down-flows in the structure.

\subsection{Density}

The \ion{Mg}{vii} ratio used in the previous section yields a density
of $\log\,N_{\rm e}$=$10.63^{+0.10}_{-0.09}$ for the ARTRB. Note that,
because of 
the non-Gaussian profiles, it was necessary to use the \ion{Si}{vii}
\lam275.35 line to estimate the \ion{Si}{vii} contribution to
\ion{Mg}{vii} \lam278.39. This high density is consistent with
the high 
densities measured by \citet{widing82}, \citet{young97} and
\citet{young04} for the transition region events they studied using
Skylab and SOHO/CDS spectra, respectively.

\subsection{Continuum emission}

\begin{figure}[h]
\centerline{\epsfxsize=8.5cm\epsfbox{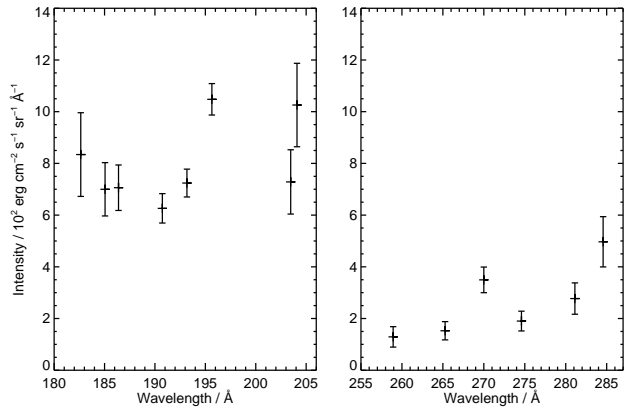}}
\caption{Plot showing continuum emission for the ARTRB in the two EIS
  wavelength bands. The continuum level has been estimated from
  spectral regions believed to be free from emission lines.}
\label{fig.accurate-continuum}
\end{figure}

Another feature demonstrated by this small brightening is enhanced
continuum emission in the short wavelength band
(Fig.~\ref{fig.accurate-continuum}).
The continuum level has been estimated from regions apparently free of
emission lines, however Fig.~\ref{fig.accurate-continuum} suggests
that some of the regions chosen are affected by emission lines.
The continuum is probably due
to recombination onto H-like helium. The photoionization edge of
He$^+$ is at 227.8~\AA, thus explaining why there is little continuum
enhancement in the long wavelength band. 
%% Categorization of the continuum
%% emission is difficult with the current data-set as there are few
%% wavelength regions that can unambiguously identified as being free of
%% emission lines. 
Note that the continuum level appears to rise at the
longest wavelengths in the 246--292~\AA\ band. This is probably the
short wavelength wing of \ion{He}{ii} \lam304, given the significant
broadening of other cool lines in the EIS spectrum
(Fig.~\ref{fig.line-profiles}).

\subsection{Time evolution}

The short lifetime of this ARTRB (as judged from the TRACE data) means
that high cadence (at most a few minutes between rasters) EIS rasters
are required to study such events in detail. EIS has the capability of
automatically switching to a new study if a threshold intensity is
reached in a specified emission line, and this will be a valuable tool
for studying such short-lived features. EIS slot movies are also an
alternative as the small spatial scale of the events should limit any
spatial--spectral ambiguity introduced by using a wide slit. XRT data
would appear to be of limited value due to the low temperature of the
events, but SOT data will be extremely important for relating the ARTRBs
to the magnetic field and photospheric/chromospheric plasma. A cursory
glance of \ion{Ca}{ii} images reveals that the ARTRB here is located
at the edge of a plage region and is bright in \ion{Ca}{ii}.

%% [Note that there are no TRACE data between 22:55 and 23:42.]

%% [There are 1600, 1700 and WL data at 22:48 and the brightening can be
%% seen in these.]

%% [SOT data seem to stop at 22:36 - this may just be missing data,
%% though.]

%% [SOT catches the region where the bright point occurs, but not the
%% footpoints.]

\section{Conclusions}

The EIS active region data-set shown here shows two types of structure
that yield strongly enhanced transition region emission
lines. Footpoints of large active region loops that are most
characteristically seen in  TRACE 171 filter images can clearly be seen
in emission lines of \ion{Mg}{v--vii}, \ion{Si}{vii} and
\ion{Fe}{viii} -- ions formed at temperatures $\log\,T=5.4$--5.8. The
density in these footpoint regions can be accurately measured using
the \ion{Mg}{vii} \lam280.75/\lam278.39 ratio, and a fall-off in
density from $3\times10^9$ to $1.5\times 10^9$~cm$^{-3}$ from the
base of a loop to a (projected) height of 20~Mm is found here.

A strong active region transition region brightening (ARTRB) is also
seen in the data-set and demonstrates very strong emission in
\ion{O}{v} \lam192.9. The ARTRB shows broadened emission line
profiles and extended wings. The \ion{Mg}{vii} \lam280.75/\lam278.39
yields a density of $4\times 10^{10}$~cm$^{-3}$. The brightening is
short-lived, but co-spatial TRACE images demonstrate a highly dynamic,
small loop-like feature with a lifetime of around 60~minutes.

The value of including transition region lines in EIS studies has been
demonstrated here, and observers are recommended to include one or
more of the emission lines listed in Table~\ref{tbl.lines} when
designing EIS studies.

~

Hinode is a Japanese mission developed and launched by ISAS/JAXA, 
with NAOJ as domestic partner and NASA and STFC (UK) as international 
partners. It is operated by these agencies in co-operation with ESA 
and NSC (Norway). The authors thank 
D.~Bewsher for useful comments on the manuscript. G.~Del Zanna thanks the
hospitality of DAMTP, University of Cambridge.

%% \emph{Hinode} is an international project supported by JAXA, NASA,
%% STFC and ESA. We are grateful to the \emph{Hinode} team for all their
%% efforts in the design, build and operation of the mission. The authors
%% thank 
%% D.~Bewsher for useful comments on the manuscript.

%% For observers looking to design EIS observation studies for active
%% regions, the emission lines listed in Table~\ref{tbl.lines} are
%% recommended

%% \section{Line IDs}

%% The line at 258.1 that comes up strong in the small brightening would seem
%% to be Si IX, which is strongly density sensitive. It may be possible
%% to make an estimate of the density using the ratio with the nearby Si
%% X line. Note that there is also the Si X diagnostic.

%% Fe XIII and Fe XI suggest a higher density in brightening than
%% surroundings, but not by a huge factor.

%% \section{Continuum image}

%% I probably need to do a careful job of estimating the pedestal and
%% dark current. Suggestion:

%% * concentrate only on the exposure containing the brightening

%% * go along the slit length to estimate the background level in each of
%%   the CCD quadrants. Basically need to look at background emission in
%%   dark areas

%%%
% See the manual for the detail.
%%%


\begin{thebibliography}{}
% Journals(e.g. A\&A,ApJ,AJ,NMRAS,PASP ...)
% Authors, Year, Journal, Vol#, Page#
% Journal Title Abbreviation >> http://www.asj.or.jp/pasj/Jabb.html
\bibitem[Culhane et al.(2007)]{culhane07}
   Culhane, J. L., Harra, L. K., James, A. M., et al. 2007,
   Solar Physics, in press
\bibitem[Del Zanna(2003)]{delzanna03a}
  Del Zanna, G. 2003,
  A\&A, 406, L5
\bibitem[Del Zanna \& Mason(2003)]{delzanna03b}
  Del Zanna, G., \& Mason, H. E.\ 2003,
  A\&A, 406, 1089
\bibitem[Dere et al.(1997)]{dere97}
   Dere, K. P., Landi, E., Mason, H. E., Monsignori-Fossi, B. F., \& 
   Young, P. R. 1997,
   A\&AS, 125, 149
\bibitem[Doyle et al.(2004)]{doyle04}
   Doyle, J. G., Madjarska, M. S., Dzif{\v c}{\'a}kov{\'a}, E., \&
   Dammasch, I. E. 2004,
   Sol. Phys., 221, 51
\bibitem[Feldman et al.(1992)]{feldman92}
  Feldman, U., Mandelbaum, P., Seely, J. L., Doschek, G. A., \&
  Gursky H.\ 1992, ApJS, 81, 387
\bibitem[Kosugi et al.(2007)]{kosugi07}
  Kosugi, T., Matsuzaki, K., Sakao, T., et al.  
  2007, Solar Physics, submitted 
\bibitem[Landi et al.(2006)]{landi06}
   Landi, E., Del Zanna, G., Young, P. R., Dere, K. P., Mason, H. E.,
   \& Landini, M. 2006,
   ApJS, 162, 261
\bibitem[Mazzotta et al.(1998)]{mazz98}
  Mazzotta, P., Mazzitelli, G., Colafrancesco, S., \& Vittorio,
  N. 1998,
  A\&AS, 133, 403
\bibitem[Widing(1982)]{widing82}
  Widing, K. G. 1982,
  ApJ, 258, 835
\bibitem[Young \& Mason(1997)]{young97}
   Young, P. R., \& Mason, H. E.\ 1997,
   Sol. Phys., 175, 523
\bibitem[Young(2004)]{young04}
   Young, P. R.\ 2004, in
   Proc. 13th SOHO Workshop, ESA SP-547, page257
\bibitem[Young(2007)]{young07}
   Young, P. R.\ 2007, in ASP Conf. Ser., 369, New Solar Physics with
   the Solar-B Mission,
   ed. K. Shibata, S. Nagata, \& T. Sakurai (San Francisco: ASP), in
   press 
\bibitem[Young et al.(2007)]{young07b}
   Young, P. R., Del Zanna, G., Mason, H. E., et al. 2007,
   PASJ, this issue
\end{thebibliography}
\end{document}